\documentclass[apj]{emulateapj}
\usepackage{apjfonts}
\shorttitle{Large-Scale Parker Winds in AGNs}
\shortauthors{Everett \& Murray}
\slugcomment{Accepted for publication in the Astrophysical Journal}

\newcommand{\oiii}{{\textsc{[O~III]}}}
\newcommand\hst{\textit{HST}}

\begin{document}

\title{Large-Scale Parker Winds in Active Galactic Nuclei}

\author{John E. Everett\altaffilmark{1} and Norman Murray} \affil{Canadian
Institute for Theoretical Astrophysics, University of Toronto, 60
Saint George Street, Toronto, ON M5S 3H8, Canada}
\email{everett@physics.wisc.edu, murray@cita.utoronto.ca}
\altaffiltext{1}{Current address: Departments of Astronomy and Physics, and Center
for Magnetic Self-Organization in Laboratory and Astrophysical
Plasmas, University of Wisconsin, Madison WI 53706 USA}

\begin{abstract}
We build and test Parker-wind models to apply to observations of
large-scale ($\sim 100$pc) outflows from Active Galactic Nuclei
(AGNs).  These models include detailed photoionization simulations,
the observed radially varying mass profile, adiabatic cooling, and
approximations for clouds dragged along in the wind and the
interaction of the wind with the circumnuclear ISM of the galaxy.  We
test this model against recent \hst/STIS observations of \oiii\
emission-line kinematics (in particular, we test against those
observed in NGC~4151, but approximately the same kinematics is
observed in NGC~1068 and Mrk~3) to constrain the viability of
large-scale thermal winds in AGNs.  We find that adiabatic cooling
dominates in these outflows, decelerating Parker winds on large
scales, making them highly unlikely as explanations of the observed
kinematics.
\end{abstract}


\section{Introduction}\label{Intro}

Outflows are ubiquitous in astrophysics: they rather paradoxically
accompany collapse processes, whether the collapse of a molecular
cloud core to form a young stellar object, the collapse of matter onto
a galaxy's central black hole, or the collapse of stars into white
dwarfs, neutron stars, or black holes.  Outflows are important to
study for their links to those basic collapse and accretion processes,
but also because understanding outflows is crucial to studying the
impact and feedback of those winds on the surrounding medium.

Outflows from AGNs are interesting for both of the reasons given
above.  While there is widespread agreement that AGNs function via
accretion from subparsec-scale disks onto super-massive black holes
\citep[see, e.g.,][]{Peterson97, Krolik99}, the detailed physical
processes controlling, for instance, the generation of the continuum,
obscuration of AGNs, transferring matter from large ($\sim 1$~kpc) to
small scales, as well as the launching mechanisms for outflows, are
still not well understood.  Most importantly here, while $\sim$50\% of
AGNs show evidence for outflowing UV and X-ray absorption
\citep{Reynolds97, George98, Crenshaw99, Kriss01}, the physical
mechanisms powering those outflows remain unclear.  Magnetic fields
are a leading candidate for launching winds in both AGNs and young
stellar objects \citep{BP82, EBS92, CL94, Bottorff97, BF00,
Bottorff00}.  In addition, given the high-flux radiation field in some
AGNs (especially the high-luminosity quasars), it is also natural and
useful to examine radiative acceleration of outflows \citep{MT75,
Shields77, Icke77, SVS85, ALB94, MCGV95, CM96, Proga00, CN01, CN03a,
CN03b, PK04}.  Some groups have also combined these two mechanisms,
addressing the importance of radiative acceleration within magnetic
winds \citep{KK94, dKB95, EKA02, E05}.  Apart from these two models,
thermal wind models have also been proposed; several papers have
examined X-ray heated winds
\citep{BMS83,KV84,BK93,KK95,Woods96,Krolik97}.  \citet{CN05} examined
the possibility of thermally-driven (Parker) winds and the ability of
those winds to explain observations of low velocity ($v \sim
500$~km~s$^{-1}$) X-ray absorption features.  In this paper, we also
examine the possibility of Parker winds, but in a different context;
we test these wind models against spatially resolved observations of
the kinematics of forbidden-emission lines (here, \oiii) in local,
low-luminosity AGNs.

Forbidden-line emission is very useful for testing models of AGN
outflows and their interaction with surrounding gas because such
emission is seen at relatively large distances from the nucleus (of
order 100~pc).  A large distance for the emitting gas is consistent
with the observed narrow (low velocity dispersion) emission lines;
this region is therefore known as the Narrow Line Region (NLR).  The
large distance to the NLR is important because whereas many of the
physical processes in AGNs occur on scales much too small to directly
observe and resolve, the NLR can be and is resolved in some AGNs.

Resolving the NLR is largely a product of the high resolution
observations of the \textit{Hubble Space Telescope}, although early
ground-based studies of NGC~4151 revealed the possibility of large
scale acceleration as early as 1990 \citep{Schulz90,MCM96,MC96}.  More
recent spectra taken with the \hst's Space Telescope Imaging
Spectrograph (STIS), however, allow high spatial and spectral
resolution studies of the NLR, offering data sufficient to constrain
photoionization and dynamical models of the outskirts of AGNs.  In
particular, \citet{Das05} have mapped the kinematics of up to three
different flux components of \oiii\ emission in NGC 4151 in five
different slits across the observed NLR.  In addition, previous
studies of the NLR of NGC 4151 \citep{Winge99, Nelson00, Kraemer00}
and of the NLR of other AGN \citep[e.g.,][]{CK00,KC00,Ruiz05} have
found similar kinematics.  The most striking result from these studies
is the apparent near-linear increases in velocity ($v \propto r$) over
scales from 10~pc to 100~pc and then an apparent linear decrease in
velocity at larger scales; further, this trend has been observed on
\textit{similar scales} in \textit{three different objects}: NGC~4151
\citep{Das05}, NGC~1068 \citep{CK00}, and Mrk~3 \citep{Ruiz01}.

These observations may help constrain the driving forces behind
large-scale AGN winds.  At first glance, both radiative acceleration
and magnetic acceleration are problematic with respect to these
observations: both mechanisms accelerate gas to its escape velocity on
length scales of order the launching radius.  If the outflow is
launched from an accretion disk, the launching scale must be $\la
1$~pc \citep{Goodman03}, and therefore both radiative and magnetic
winds would be expected to reach their terminal velocities by
approximately 10~pc from the central source.  Another possibility for
the driving is thermal, or Parker, winds: could such thermal winds
accelerate gas slowly enough to explain the observations?  This seems
plausible, especially if we allow the \oiii\ emission to originate in
clouds dragged along with the wind, which would further slow the
observed acceleration.  We can then ask what constraints these
observations of $v \propto r$ outflows place on thermal wind models.
Also, just as important: in the three cases mentioned above, what
decelerates the gas?  Finally, such models may help constrain the
geometry of the NLR \citep{Das05}.

To address these questions we must build detailed kinematic and
photoionization models of the core of AGNs and their winds.  In this
paper, we build such a thermal wind model (in
\S\ref{ParkerWindModel}), adapted to considerations of large-scale
flows in AGNs, and then apply that model to \hst/STIS observations of
NGC~4151.  The particular spectral energy distribution and $M(r)$
profile in NGC~4151 are introduced in \S\ref{NGC4151Inputs}.  The
particular mass profile that we adopt allows for a good first-order
approximation to the wind as roughly isothermal, so isothermal wind
models are applied to NLR observations of NGC~4151 in
\S\ref{isothermal}.  Isothermality in AGN winds is then tested in the
particular case of these thermal winds (and tested against variations
in the input spectrum) in \S\ref{testingIsothermality}.  We find that
thermal wind models face very stringent constraints that make it very
unlikely that they can explain the observed NLR kinematics, due
chiefly to the effects of adiabatic cooling which ensures that the
winds are not isothermal, and in fact cool on large scales ($r \ga
5$~pc).

\section{Parker Winds}\label{ParkerWindModel}

We first review the basic Parker (solar) wind model
(\S\ref{OverviewParkerWind}), and then introduce changes to that model
that must be included when hypothesizing a large-scale Parker wind at
the center of an active galaxy: the increase in enclosed mass as a
function of radius, adiabatic cooling for outflows at such large
distances, and finally the effects of wind/cloud and wind/ISM drag.

\subsection{Parker Winds Applied to AGNs}\label{OverviewParkerWind}

A Parker wind can be thought of as an extended thermal wind.  To
explain this terminology, first consider the simplest picture of a
thermal wind, where gas at its launch point is heated to high
temperature such that that its sound speed is greater than the escape
speed ($c_s > v_{\rm escape}$).  This requirement can be recast to ask
if the gas temperature is greater than the ``escape temperature'' $T_g
\equiv GM\mu/Rk$ \citep{BMS83}, where $M$ is the mass of the central
object, $\mu$ is the mean mass per particle, $R$ is the launching
radius, and $k$ is Boltzmann's constant.  An inherent difficulty with
this picture, in relation to the NLR data discussed in the
Introduction, is that there is no extended acceleration as is
apparently observed.

A Parker wind also depends on thermal effects, but with the major
difference that the gas is not heated at the base of the wind to have
$c_s > v_{\rm escape}$.  The wind is therefore more like an extended
atmosphere, where, at the base of the wind, $c_s \ll v_{\rm escape}$.
The key difference, though, between a Parker wind and an atmosphere is
that if such strongly bound gas is heated as a function of height such
that the temperature drops more slowly than $M(r)/r$, where $M(r)$ is
the enclosed mass and $r$ is the spherical radius, the atmosphere must
expand supersonically \citep[explained in more detail later in this
section; see][]{Parker65}.  This leads to rather slow acceleration,
compared, for instance, to radiatively-driven and magnetically-driven
winds.

As mentioned previously, therefore, this model is therefore appealing
for the NLR of AGNs.  This is especially true of NGC~4151, as
photoionization analyses have shown that the central AGN continuum
dominates the heating of the NLR \citep[][and references
therein]{Alexander99,Nelson00,Kraemer00,Kraemer01}.  These
photoionization analyses also show that shock heating \citep[which
could, for example, result from an interaction with a collimating
medium, e.g.][]{BM06} is also not likely to be important.  (We note,
though, that Kraemer et al. 2001 find the possibility of shock heating
near the start of deceleration of the wind at approximately $r \sim
100$~pc.)  It is also possible in some AGNs for the central radio jet
to affect the NLR by increasing the NLR line widths
\citep[e.g.,][]{NW96, Capetti97, Steffen97, Axon98} but that does not
appear to be the case for NGC~4151 \citep{Das05}.

To examine this possibility in more detail, we start with the basic
equation for an isothermal Parker wind:
\begin{equation}
\frac{dv}{dr} = \frac{\frac{2 c_s^2}{r} - \frac{GM}{r^2}}{v \left(1 -
\frac{c_s^2}{v^2} \right) }\label{dvdrEq}
\end{equation}
where $v$ is the one-dimensional radial velocity, and $r$ is the
spherical radius.

\begin{figure}[h]
\begin{center}
\includegraphics[angle=-90, width=8.5cm]{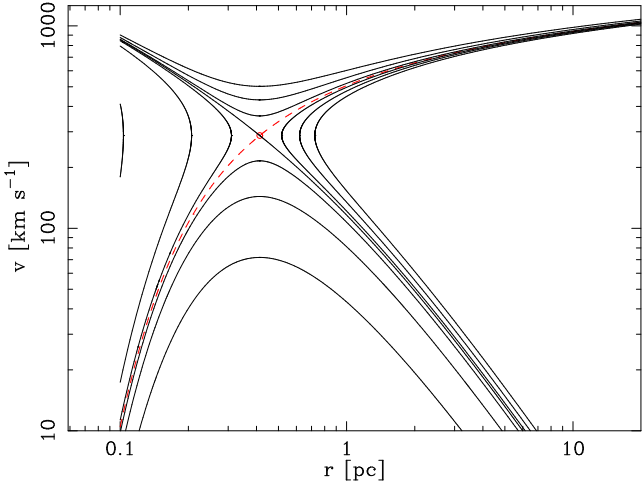}
\caption{A variety of solutions to the Parker wind equation
  (Eq.~\ref{dvdrEq}) for an isothermal wind with $T = 5 \times
  10^6$~K, $M_{\bullet} = 1.3 \times 10^7~M_{\odot}$, and $r_{\rm
  launch} = 0.1$~pc.  The dashed line shows the solution that passes
  from low velocity at the launch point, through the critical point,
  to high velocity at large distance.  This solution, passing through
  the critical point, is the only steady-state solution that yields a
  viable wind model.  Other solutions, not passing through the
  critical point, are shown for comparison.  For completeness, we also
  show that it is possible to have an inflow solution that passes
  through the critical point: gas at high velocity ($v \gg c_s$) at
  large distance ($r \gg r_{\rm critical}$) that slows as it falls to
  the central object; this is known as Bondi-Hoyle accretion
  \citep{BH44}.\label{parkerWindSolns}}
\end{center}
\end{figure}

Two important insights can be seen from examining
Equation~\ref{dvdrEq} closely.  First, we note the importance of the
``critical point'' in this equation.  As stated earlier, gas at the
base of a Parker wind is gravitationally bound, so that $GM/r^2 \gg
c_s^2/r$.  Therefore, the numerator of Equation~\ref{dvdrEq} must be
negative at the base of the wind; since the denominator is also
negative ($v < c_s$), a positive acceleration results.  However, as
the wind moves outward and $v$ increases, eventually $v = c_s$ and the
denominator of Equation~\ref{dvdrEq} becomes zero.  In order for the
Parker wind equation to be physically meaningful, the numerator must
also be equal to zero at the same time the denominator becomes zero,
so $2 c_s^2/r$ must equal $GM/r^2$ (or $r = r_{\rm critical} =
GM/2c_s^2$) must be true when $v = c_s$.  This requirement sets the
location of the critical point: this is the ($r$,$v$) pair that the
solution must ``thread'' so that the wind solution can describe a flow
that continuously accelerates.  In fact, the solution that passes
through this critical point is the only solution that allows an
outflow to large distances and to speeds above the escape velocity.
Physically, this critical point represents the last point downstream
in the outflow where the boundary conditions at the disk can
communicate with and therefore affect the wind.  An example of such a
solution is shown by the dashed line in Figure~\ref{parkerWindSolns},
which threads the critical point, indicated by the circle.  Other
solutions (such as the ``breeze'' solutions that have $dv/dr = 0$
below the position of the critical point in
Fig.~\ref{parkerWindSolns}, achieving smaller peak velocities than the
critical point velocity and then decelerating) that do not result in a
viable outflow are also shown.

Second, in order to pass through that critical point, there is a
further constraint on the temperature structure of the outflow, which
was mentioned briefly above.  In order for the solution to pass
through the critical point and continuously accelerate, the
temperature profile, $T(r)$ must decrease with radius less quickly
than $M(r)/r$.  To show this, we recast the Parker wind equation into
a form that shows directly the comparison between the pressure
gradient and gravity:
\begin{equation}
v \frac{dv}{dr} \left(1 - \frac{c_s^2}{v^2} \right) = -r^2
\frac{d}{dr} \left( \frac{c_s^2}{r^2} \right) - \frac{GM}{r^2}\label{compareEq}
\end{equation}
This equation is simply a more general form (not assuming
isothermality) of Equation~\ref{dvdrEq} which will allow us to
investigate non-isothermal wind models.  The same conditions that
applied to the numerator of Equation~\ref{dvdrEq} apply to the
right-hand side of Equation~\ref{compareEq}: namely, it will also have
to transition from negative to positive values at the critical point.

Now, instead of assuming an isothermal wind (as in Eq.~\ref{dvdrEq}),
here we set $T = T(r) = T_0(r/r_0)^\alpha$, but retain the assumption
of a central point mass, $M$.  After substituting this into
Equation~\ref{compareEq} and simplifying, we find:
\begin{equation}
v \frac{dv}{dr} \left(1 - \frac{c_s^2}{v^2} \right) = -
\frac{c_{s,0}^2}{r_0^\alpha} (\alpha-2) r^{\alpha-1} -
\frac{GM}{r^2}\label{alphaCompare}
\end{equation}
where $c_{s,0}$ is the sound speed at the base of the wind.  One can
see that in order for the pressure term to dominate gravity, $\alpha$
must be greater than -1.  To illustrate this, consider the case of
$\alpha = -1$ in Equation~\ref{alphaCompare}:
\begin{equation}
v \frac{dv}{dr} \left(1 - \frac{c_s^2}{v^2} \right) = \frac{3
c_{s,0}^2 r_0}{r^2} - \frac{GM}{r^2}
\end{equation}
At the base of the wind, the right-hand side of the equation will read
$3 c_{s,0}^2/r_0 - GM/r_0^2$.  Since, at the base of a Parker wind,
$c_{s,0}^2 \ll GM/r_0$, this term will be negative at the base of the
wind, and since both the pressure gradient term (the first term on the
right-hand side) and the gravitational term decrease as $r^2$, the
right-hand side of the equation will never be positive.  In that case,
the wind will not accelerate to escape velocity.  Therefore, in order
for the wind to accelerate in this case (where $M$ is constant with
radius), $\alpha$ must be greater then -1.  One can derive the same
result another way: if $\alpha \leq -1$, the result is a static
atmosphere \citep[see][]{Parker65}.

For large-scale winds ($r \ga 1$~pc) in AGNs, though, $M(r)$ increases
with radius.  Including the possibility of a spatially-varying $M(r)$
in our requirement on the temperature profile $T(r)$, above,
$r^\alpha$ must decrease less quickly than $M(r)/r$ in order to
accelerate the wind to large distances.  Assuming an isothermal sphere
to model the gravitational potential of the circumnuclear matter, we
will find that (for NGC~4151) the black hole stops dominating the
gravitational potential at approximately 1~pc.  Beyond that distance,
$M(r) \propto r$.  In this regime, assuming a simple $M(r) =
M_0(r/r_0)$ for now, Equation~\ref{alphaCompare} becomes:
\begin{equation}
v \frac{dv}{dr} \left(1 - \frac{c_s^2}{v^2} \right) = -
\frac{c_{s,0}^2}{r_0^\alpha r} (\alpha-2) r^{\alpha} -
\frac{GM_0}{r_0 r}\label{varyMCompare}
\end{equation}
Equation~\ref{varyMCompare} then shows that we must have $\alpha > 0$
in order to get large-scale acceleration in a thermal wind.  In other
words, the wind must have $T(r)$ at least slightly increasing with
radius in order for a Parker wind to accelerate.

In order to check this constraint on $T(r)$, we must ensure that we
address all of the heating and cooling components in an AGN wind
accurately.  In \S\ref{sed}, we will use Cloudy \citep[Version
05.07.06;][]{Ferland98} to simulate the photoionization heating,
although Cloudy by itself cannot include the effects of adiabatic
cooling in the context of our thermal wind and wind geometry (it can
calculate adiabatic cooling in the context of its own internal wind
models, however).  This important cooling source will be considered
next.

\subsection{Adiabatic Cooling}\label{adiabatic}

We will run photoionization simulations at regular positions along the
flow, but those separate photoionization simulations will not have any
knowledge about the kinematics of the outflow.  This would problematic
for these large-scale flows where adiabatic cooling is important
\citep[see, e.g.][]{RGS90,Frank92,Safier93,Shang02,CN05}, so we
calculate the magnitude of the adiabatic cooling outside of Cloudy and
include it as input to Cloudy.  The adiabatic cooling is simply
calculated from:
\begin{equation}
\frac{dq}{dt} = \frac{d\epsilon}{dt} + P\frac{d}{dt} \left( \frac{1}{\rho} \right) 
\end{equation}
where $q$ is the heat per unit mass, $\epsilon$ is the internal energy
per unit mass, and $P~d(1/\rho)$ is the work per unit mass done by
the expanding gas.  After some algebra, the change in internal energy
is given by:
\begin{equation}
\rho \frac{d}{dt} \left( \frac{3}{2} \frac{P}{\rho} \right) =
\frac{P}{\rho} \frac{d\rho}{dt}
\end{equation}
Finally, using $\dot{M}_{\rm out} = 4 \pi r^2 \rho v$ to eliminate
$\rho$ from the above expression, we find that the adiabatic cooling
is given by:
\begin{equation}
\Lambda_{\rm adiabatic} = \rho c^2_{\rm s} \left(\frac{2v}{r} +
\frac{a}{v}  \right) {\rm ergs~cm^{-3}~s^{-1}}
\end{equation}
where $\rho$ is the mass density of the outflow, $c_s$ is the speed of
sound, $v$ is the radial velocity, $r$ is the distance from the
central source, and $a$ is the acceleration in the wind.  This
particular expression and the above-noted units are those necessary to
supply the correct cooling rate to Cloudy; as such, the units on this
expression are different from, e.g., \citet{DM72}.

For all of the wind solutions found in this work, $a/v$ is at least
three orders of magnitude smaller than $2v/r$, so the $a/v$ term is
safely neglected.

\subsection{Clouds Dragged Via Ram Pressure}

We will model the \oiii\ emission as coming from clouds injected into
the wind and accelerated by the ram pressure of the wind on those
clouds.  The drag force for each individual cloud is computed via
\citep[e.g.,][]{Smith84}:
\begin{equation}
F_{\rm drag, cloud} = \rho_{\rm wind} (v_{\rm wind} - v_{\rm cloud})^2
A_{\rm cloud}
\end{equation}
where $\rho_{\rm wind}$ is the mass density of the wind, $v_{\rm
wind}$ and $v_{\rm cloud}$ denotes the velocity of the wind and
clouds, and $A_{\rm cloud}$ represents the cross sectional area of a
cloud.  Assuming a spherical cloud with mass $M_{\rm cloud} = (4/3)
\pi r^3 \rho_{\rm cloud}$, the acceleration for each cloud is
\begin{equation}
a_{\rm drag, cloud} = \frac{\rho_{\rm wind}}{\rho_{\rm cloud}} 
  \frac{\frac{3}{4}(v_{\rm wind} - v_{\rm cloud})^2}{R_{\rm cloud}} = 
\frac{n_{\rm wind}}{N_{\rm H,cloud}} 
  \frac{3}{4}(v_{\rm wind} - v_{\rm cloud})^2\label{cloudDragEq}
\end{equation}
where $R_{\rm cloud}$ is the radius of an individual cloud.  The ratio
$n_{\rm wind}/N_{\rm H,cloud}$ is a free parameter in our models.  We
use this parameter for all clouds, although it would be more
physically plausible to imagine a range of densities for the clouds
intercepting the large-scale thermal wind.  This would lead to a range
of velocities, which might explain the large ($\sim 500$~km~s$^{-1}$)
\oiii\ line widths observed in this object \citep{Das05} and seen in
many AGNs \citep{Rice06}.  The model presented here solely addresses
trends in the observed centroid of the lines, however, and so only a
single density ratio is used.

We assume that the drag of the clouds on the wind is negligible
compared to the acceleration of the Parker wind.

\subsection{Decelerating the Parker wind via ISM drag}

The wind is decelerated in much the same way that the clouds are
accelerated: ram pressure from the impact on an external medium, which
we assume is also broken up into clouds, and is also assumed
stationary.  Much like before, then, the drag force is given by:
\begin{equation}
a_{\rm drag, ISM~clouds} = \frac{\rho_{\rm wind}}{\rho_{\rm ISM~cloud}} 
  \frac{\frac{3}{4}v_{\rm wind}^2}{R_{\rm ISM~cloud}} = \frac{n_{\rm
  wind}}{N_{\rm H,ISM~cloud}} \frac{3}{4}v_{\rm wind}^2
\label{windDragEq}
\end{equation}

As the wind decelerates, the embedded clouds decelerate as well, as
$v_{\rm wind}$ becomes smaller than $v_{\rm cloud}$ in the cloud drag
equation, Eq.\ref{cloudDragEq}.  Again, $n_{\rm wind}/N_{\rm
H,ISM~cloud}$ is a free parameters in the models we run.

The wind decelerates under this drag force until the wind becomes
subsonic, at which point the integration of the equations of motion
ceases.  However, $v \la c_s$ at such large distances that it does not
affect our ability to fit the observed kinematics.

\section{Applying the Parker Wind Model to NGC~4151}\label{NGC4151Inputs}

The discussion in \S\ref{ParkerWindModel} outlined the basic physics
of the Parker wind.  Next, the inputs to that model must be defined
from the observations of, in this case, NGC~4151.  Most importantly
for the consideration of heating, the central spectral energy
distribution must be realistically modeled (\S\ref{sed}) and the
particular $M(r)$ profile for NGC~4151 must be included
(\S\ref{Mvsr}).

\subsection{Modeling the Central Spectral Energy Distribution}\label{sed}

\begin{figure}[h]
\begin{center}
\includegraphics[angle=-90, width=8cm]{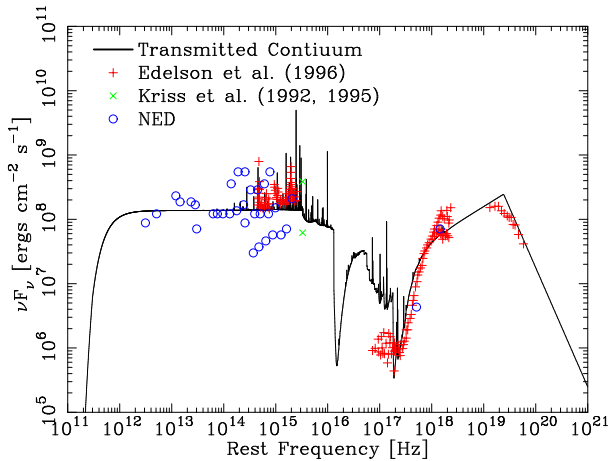}
\caption{Comparison of the assumed central SED for NGC~4151 (defined
  in Table~\ref{sedParams} with added intrinsic absorption) with
  observations.  The solid line shows the SED, modeled by Cloudy,
  after absorption by components ``X-High'' and ``DEa'' from
  \citet{Kraemer00}.  The red data points show the optical, UV, X-ray
  and $\gamma$-ray results from \citet{Edelson96}, the green data
  points show the range in UV brightness observed during \textit{HUT}
  pointings \citep{Kriss92, Kriss96}, and the blue data points show
  the wide array of spectral data in NASA's Extragalactic
  Database.\label{sedCompare}}
\end{center}
\end{figure}

An accurate central spectral energy distribution (SED) is very
important to understanding the heating and subsequent acceleration of
Parker winds in NGC~4151.  To model the SED, we use the central
continuum of this object as modeled by \citet{Kraemer00}; this
continuum is then used as input for the Cloudy photoionization
simulations.  The continuum has two components: an intrinsic SED and
absorption from surrounding gas.  The specification for the intrinsic,
unabsorbed SED is given in Table~\ref{sedParams}: the first and last
values of $\alpha$ in the table are relatively unimportant, and are
specified simply to cut the spectrum off at low and high energies,
while the intermediate values of $\alpha$ are from \citet{Kraemer00}.
To this simple power law continuum, we add absorption via an X-ray
warm absorber and UV absorber \citep[components ``X-High'' and ``DEa''
from][]{Kraemer00}.  The ``X-High'' absorber has a total hydrogen
column density of $N_{\rm H} = 10^{22.5}$~cm$^{-2}$ and an ionization
parameter of $U = n_{\rm ion}/n_{\rm H} = L_{\rm ion}/4 \pi r^2 n_{\rm
H} c = 1.0$ (where $L_{\rm ion}$ is the ionizing luminosity for $E >
13.6$~eV), while the ``DEa'' absorber has a total hydrogen column of
$N_{\rm H} = 10^{19.75}$~cm$^{-2}$ with $U = 10^{-3}$ \citep[using the
column for the ``northeast'' absorber from][]{Kraemer00}; the
metallicities listed in \citet{Kraemer00} are used for these
absorption systems and therefore used for all of the photoionization
simulations presented here for self-consistency.  (The metallicities
from \citet{Kraemer00} includes only those metals with abundances
greater than $\sim 10^{-5}$ of hydrogen, and for those metals, differs
only by 0.2 dex from Cloudy's default solar metallicity.  Using
Cloudy's default solar metallicity for the remaining calculations
results in only insignificant differences in the $T(r)$ calculations
in \S\ref{photoResults}.) The absolute luminosity was set to
$10^{43.56}$~erg~s$^{-1}$ in the range of 13.6~eV to 1~keV to match
the spectrum displayed in Figure 2 of \citet{Kraemer00}.  The
resultant continuum is then compared to a range of observations in
Figure~\ref{sedCompare}.  The generally good agreement between the
model continuum and the observations is essential for accurately
modeling thermal winds in this source.

It is important to note, of course, that the luminosity of NGC~4151
has been observed to vary by factors of a few over the past decade
\citep[see, e.g.,][where the continuum level was seen to be roughly
2.5 times brighter in the 2-10 keV band in 2002 vs. observations in
2000]{Kraemer05}.  The continuum that we have assumed here fits
reasonably well to the observations of \citet{Edelson96}, taken in
December 1993 when NGC~4151 was near its peak historical brightness,
so tests with this continuum should yield good estimates for maximum
heating of thermal winds.  We will also check the dependence of the
results on the central continuum in \S\ref{altContinuum}.

\begin{deluxetable}{lll}
\tablecaption{Parameters for NGC~4151's intrinsic, unabsorbed SED\label{sedParams}} 
\tablehead{
\colhead{$E_{\rm begin}$} & \colhead{$E_{\rm end}$} & \colhead {Power
  Law $\alpha$: $F_{\nu} \propto \nu^{\alpha}$}}
\startdata
1.36 $\times 10^{-7}$ eV & 1.24 $\times 10^{-3}$ eV & 2.5 \\
1.24 $\times 10^{-3}$ eV  & 13.6 eV & -1.0 \\
13.6 eV & 1 keV & -1.4 \\
1 keV & 100 keV & -0.5 \\
100 keV & 100 MeV & -2.86 \\
\enddata
\end{deluxetable}

\begin{figure}[h]
\begin{center}
\includegraphics[angle=-90, width=8cm]{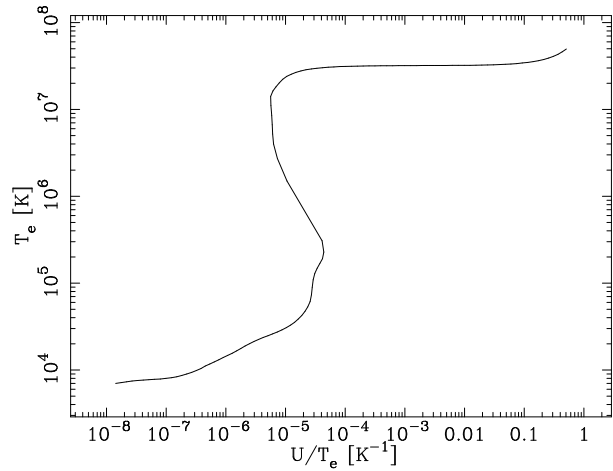}
\caption{The equilibrium ``S'' curve for the central absorbed SED in
  Figure~\ref{sedCompare}.  This plot shows that the Compton
  temperature of gas near the central source in NGC~4151 is $3.3
  \times 10^7$~K.  Note the upturn in $T_{\rm e}$ at $U/T_{\rm e} \ga
  0.1$ due to stimulated Compton scattering.\label{comptonTemp}}
\end{center}
\end{figure}

\begin{figure}[h]
\begin{center}
\includegraphics[angle=-90, width=8cm]{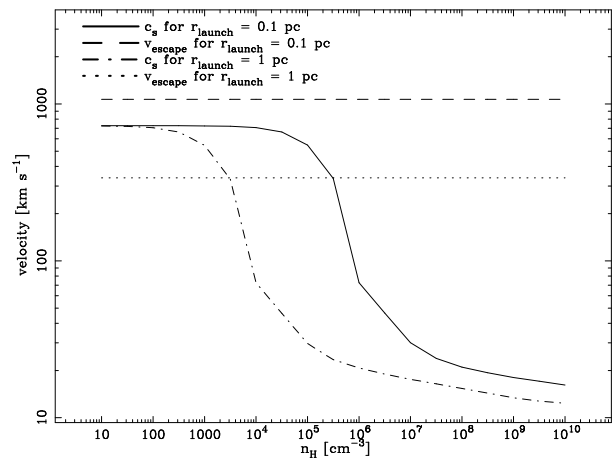}
\caption{The sound speed, $c_{\rm s}$, vs $n_{\rm H}$, the number
  density, at the disk, using the SED shown in Fig.~\ref{sedCompare}.
  Only densities for which $c_{\rm s} > v_{\rm escape}$ can gas be
  ``evaporated'' off of the disk.  This figure shows that such
  evaporation only happens for large distances ($r \ga 1$~pc) and very
  small densities ($n_{\rm H} \la 3 \times 10^3$~cm$^{-3}$).  This gas
  is not appreciably accelerated at large distances as is observed, so
  this model cannot explain the observed velocities in NGC~4151, but
  it does hint that more detailed thermal winds might be viable
  models, since approximately the correct velocities can be achieved.
  \label{simpleThermalWindTest}}
\end{center}
\end{figure}

We next calculate the Compton temperature for this gas, using the
above SED and luminosity to calculate the gas temperature at a range
of values for the gas ionization parameter, $U$.  The resultant plot
of equilibrium gas temperatures \citep[as in][]{KMT81} is shown
Figure~\ref{comptonTemp}, which also shows that $T_{\rm e,Compton} =
3.3 \times 10^7$~K.  Note, however, that at extremely high ionization
parameters and temperatures ($U/T_{\rm e} \ga 0.1$; $U \geq 2.5 \times
10^6$ and $T_{\rm e} \geq 3.3 \times 10^7$~K), the equilibrium
temperature starts increasing again.  This is due to stimulated
Compton heating of the gas, which occurs only at very high ionization
parameters: when the radiation field is very intense compared to the
gas density (when the electrons are only an ``impurity'' in the flux
of photons), the stochastic electric field of the high-intensity
photon field modifies the electrons' response to the photons, yielding
higher heating rates \citep{LS70, B73, Wilson82}.  We include this
regime in the equilibrium curves shown here for completeness, but note
that even without adiabatic cooling, the ionization parameters in the
Parker winds explored here never reach the ionization levels where
this extreme heating dominates.  In addition, including adiabatic
cooling results in significant decreases in this maximum temperature
\citep[as already noted, for instance, by][]{CN05}.
Given the SED and equilibrium curve, we can ask if the simplest
thermal winds could explain the NLR velocities in NGC~4151.  For a
range of initial wind densities, we show the thermal speed and escape
speeds for winds launched from 0.1 and 1 parsec in
Figure~\ref{simpleThermalWindTest}.  For winds launched at 0.1~pc, the
gas is not heated to sufficient temperatures to escape the
gravitational potential.  Winds launched at $r \ga 1$~pc have sound
speeds high enough that gas can be evaporated off the disk and escape
the gravitational potential to approximately the correct velocities,
although such a simple evaporation model would not explain the
large-scale acceleration seen in NGC~4151.  However,
Figure~\ref{simpleThermalWindTest} does show that high density winds
will have $c_{s,0} \ll v_{\rm escape}$, so that Parker winds may be
viable solutions (see \S\ref{OverviewParkerWind}).
\subsection{Radially Increasing $M(r)$}\label{Mvsr}

As mentioned in \S\ref{OverviewParkerWind}, the increase in enclosed
mass with radius is crucial to modeling thermal winds in AGNs
\citep[and has been considered without photoionization simulations
by][]{KV84}.  Observationally, the mass distribution within NGC~4151
is not well understood: the only data that exists are estimates of the
black hole mass, and estimates of the galaxy's potential at
approximately 700 pc.  As a first order approximation, we assume an
isothermal sphere mass profile and so therefore connect the inferred
masses with the following functional form:
\begin{equation}
M(r) = M_{\bullet} + \frac{2r\sigma^2}{G}\label{MReq}
\end{equation}
where $\sigma(r)$ is the observed velocity dispersion at radius $r$
and $M(r)$ is the enclosed mass within radius $r$.

The central black hole mass is $M_{\bullet} = (1.33~\pm~0.46)~\times
10^{7}~M_{\odot}$ \citep{Peterson04}.  This mass must dominate at
small radii (with this $M(r)$ profile, at $r \la 2$~pc).  At large
radii, the enclosed mass can be derived from \textsc{HI} 21cm line
studies; we find the best current constraint to be $\sigma(r = 10'' =
690 {\rm pc}) = 122~$km~s$^{-1}$ \citep{Mundell99}.  These two data
points define our enclosed mass function, $M(r)$.

As discussed in \S\ref{OverviewParkerWind}, $M(r) \propto r$ at large
scales in AGN is very important for Parker wind models, as it sets
stringent constraints on the $T(r)$ profile required for the Parker
wind to be launched to infinity.  This $M(r)$ profile also changes the
position of the critical point in the wind.  Inserting
Equation~\ref{MReq} for $M(r)$ into the Parker wind equation
(Eq.\ref{dvdrEq}) and solving for the critical point condition, we
find
\begin{equation}
r_{\rm critical} = \frac{G M_{\rm BH}}{2(c_{\rm s}^2 - \sigma^2)}
\end{equation}
In the limit where $M_{\rm BH}$ is dominant ($\sigma = 0$), this
equation reduces as expected to the standard expression for the
critical point, $r_{\rm critical} = G M_{\rm BH}/2c_{\rm s}^2$.
Further, the acceleration at the critical point is also modified from
the standard Parker-wind expression because of the inclusion of the
$M(r)$ function, and is given by
\begin{equation}
\left( \frac{dv}{dr} \right)^2_{r = r_{\rm critical}} =
\frac{(c_{\rm s}^2 - \sigma^2)^3}{(GM_{\rm BH})^2} 
\end{equation}

These equations allow for integration of the wind equations starting
from the critical point.

\section{Isothermal Parker Wind Models}\label{isothermal}

\begin{figure}[h]
\begin{center}
\includegraphics[width=16cm, width=8cm]{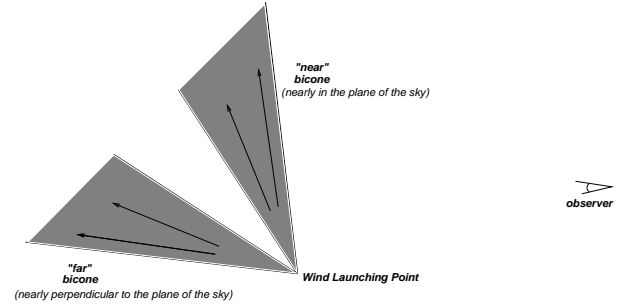}
\caption{A diagram of the biconical geometry for the observed NLR outflow in
NGC~4151.  The observer, at right, sees the outflow tilted at
an inclination angle of $45^\circ$ to the line of sight.  This yields
to halves to the bicone with somewhat distinct kinematics: a ``near''
bicone that is nearly in the plane of the sky, and a ``far'' bicone
with is closer to being perpendicular to the plane of the sky, or
nearly along the line of sight of the observer.  \label{biconeGeometry}}
\end{center}
\end{figure}

To test the Parker wind model against observations, we first
numerically solve the isothermal Parker wind model, integrating from
the wind's critical point to the disk and then from the critical point
out to where the wind decelerates to $v < c_s$.  That 1D solution to
the Parker wind is then mapped to a biconical outflow (with
geometrical parameters largely similar to those in Das et al.) and
then compared to the kinematic data.  Our geometry is depicted in
Figure~\ref{biconeGeometry}; the ``near'' component of the bicone, for
NGC~4151, is close to being in the plane of the sky, while the ``far''
component of the bicone is close to being out of the plane of the sky.
We will use this terminology in refering to the different kinematic
components of the bicone that are used to explain the spread of
observed velocities in our dynamical models.  


\begin{figure}[h]
\begin{center}
\includegraphics[angle=-90, width=8cm]{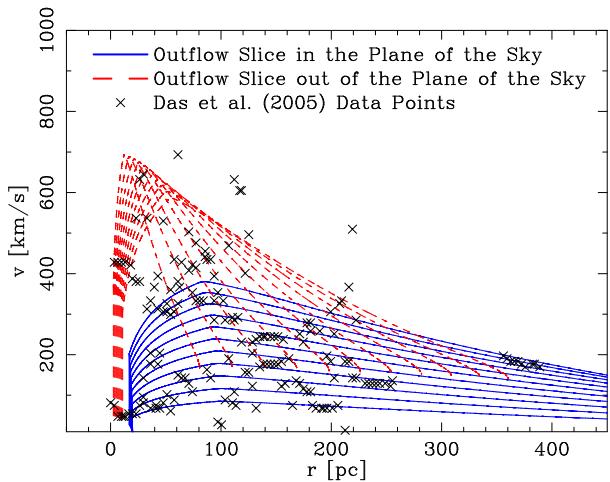}
\caption{The best-fit velocity vs radius profiles for clouds embedded
  in a biconical, \textit{isothermal} Parker wind (fit parameters are
  given in Table~\ref{cloudResults}).  The blue lines show the
  velocity results for the clouds on the near side of the outflow,
  where the outflow bicone is nearly in the plane of the sky, or
  perpendicular to the line of sight.  The red lines give the velocity
  for the component of the outflow bicone that is close to lying along
  the line of sight.  The range of lines of each color track the
  velocities for intermediate outflow angles within the assumed
  bicone.  The crosses give the observed velocities for the high-,
  intermediate-, and lowest-level \oiii\ flux emission lines from
  \citet{Das05}. \label{cloudModelBestFit}}
\end{center}
\end{figure}

\begin{figure}[h]
\begin{center}
\includegraphics[angle=-90, width=8cm]{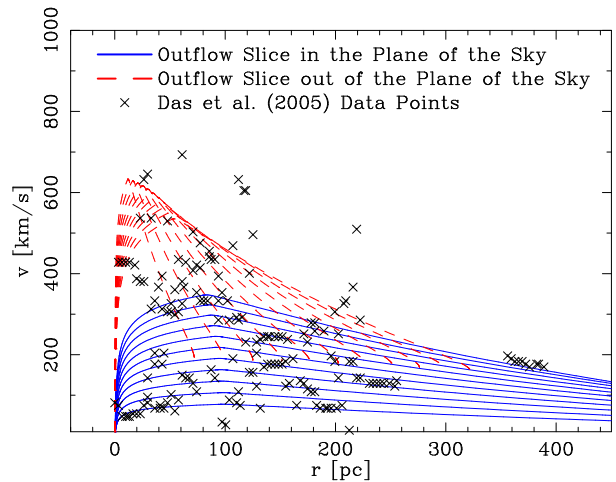}
\caption{As in Fig.~\ref{cloudModelBestFit} but velocities are those
  for the wind and not the clouds embedded in the wind.  (Note the
  different vertical velocity scale on this plot vs. the velocity
  scale in
  Figure~\ref{cloudModelBestFit}.)\label{windMotionForCloudModelBestFit}}
\end{center}
\end{figure}

Given these assumptions about the outflow geometry, the best-fit
solution to the data \citep[only for the ``north-east'' outflow in
Slit 1 of the data in][as we adopt the geometry from Das et al. and do
not fit data from other slits]{Das05} for the case of clouds being
dragged in the wind is shown in Figure~\ref{cloudModelBestFit}, where
the velocities displayed are the velocities of the embedded clouds as
a function of radius along the biconical outflow.  This model does a
reasonably good job of encompassing the observed velocity data points.
For completeness, the thermal wind velocities (the velocities in the
continuous wind itself and not the embedded clouds shown in
Fig.~\ref{cloudModelBestFit}) are displayed in
Figure~\ref{windMotionForCloudModelBestFit}.  The parameters for this
fit as well as ``error bars'' are given in Table~\ref{cloudResults}.
The most important parameter from this table and from the isothermal
wind models is the required temperature, $T_{\rm e} = 3 \times
10^6$~K.

Note that the error bars in Table~\ref{cloudResults} are only very
approximate quantities that result from manually varying each
parameter individually, and inspecting a range of fits to the data and
picking those fits and ranges of parameter values that seem to best
represent the data.  We are aware of no method to calculate a
$\chi^2$-like figure-of-merit to quantitatively analyze such data, but
present these approximate error bars to communicate the sensitivity of
the fit to various parameters.  Most importantly, these error bars
show that the isothermal wind temperature is fairly well constrained,
given the other parameters in the fit.  Also, note that the initial
radius is very much unconstrained; the only constraint we have here is
that $r_0 < 0.8$~pc, which comes from the requirement that $r_0 <
r_{\rm crit}$ for the chosen wind temperature.  The geometry of the
outflow is also fairly unconstrained in our fits to the data from one
slit; assuming that the picture of \oiii-emitting clouds is correct,
the inner half-angle of the bicone could be as small as zero, and the
outer half-angle of the bicone could be as large as $45^\circ$.  We
have chosen to retain a bicone geometry largely similar to that found
in the \citet{Das05} fits to all of their slits, although the outer
bicone half-angle adopted here is $38^\circ$, somewhat larger than Das
et al.'s estimate of $33^\circ \pm 2^\circ$.  For further comparison
to the values in Table~\ref{cloudResults}, Das et al. find a bicone
inclination of $45^\circ \pm 5^\circ$, and an inner bicone half-angle
of $15^\circ \pm 2^\circ$.

\begin{deluxetable*}{lll}
\tablecaption{Inferred Parameters for the NLR in NGC 4151 Using
  Wind/Cloud Model\label{cloudResults}} 
\tablehead{
\colhead{Quantity} & \colhead{Value} & \colhead {Range Permissible for
Reasonable Fit}}
\startdata
Wind Temperature, $T$                &  $3 \times 10^6$~K & $2.5
\times 10^6 - 3.5 \times 10^6$ \\
Launch Radius, $r_{\rm launch}$      &  $0.1$~pc &  $< 0.8$~pc \\
ISM Gas Radius, $r_{\rm ISM}$        &  $100$~pc & $50 - 150$~pc \\
$\frac{n_{\rm wind}}{N_{\rm H,ISM~cloud}}$ & $1.1
\times 10^{-21}$~cm$^{-1}$ & $8.1 \times 10^{-22} - 1.6 \times 10^{-21}$~cm$^{-1}$\\
$\frac{n_{\rm wind}}{N_{\rm H,cloud}}$ & $6.5 \times
  10^{-21}$~cm$^{-1}$ & $2.2 \times 10^{-21} - 3.2
  \times 10^{-20}$~cm$^{-1}$ \\ 
Angle of Inclination of Outflow Cone & $45^{\circ}$ & \\
Inner Cone Opening Angle             & $12^{\circ}$ & \\
Outer Cone Opening Angle             & $38^{\circ}$ & \\
\enddata

\end{deluxetable*}
Using these best-fit values for the wind, we derive a mass outflow
rate in the wind (for the case where $n_0 = 2 \times 10^9$~cm$^{-3}$)
of $\dot{M}_{\rm wind} = 6.6~C_{\rm f,wind}~M_{\odot}$~yr$^{-1}$ where
$C_{\rm f,wind}$ is the global covering fraction of the wind.  For the
wind parameters found above, $C_{\rm f,wind} = 0.2$, implying
$\dot{M}_{\rm wind} = 1.3~M_{\odot}$~yr$^{-1}$.

As mentioned previously and as indicated by the organization of
Table~\ref{cloudResults}, we can constrain only the ratios
$\frac{n_{\rm wind}}{N_{\rm H,cloud}}$ and $\frac{n_{\rm wind}}{N_{\rm
H,ISM~cloud}}$.  In addition, $T$ and $r_{\rm launch}$ are degenerate
as well; a reasonable fit can be found at larger radii if the
temperature drops commensurately.

\subsection{The Role of Rotation}

We do not include the effects of rotation in this model.  Rotation in
the Parker Wind would be important very close to the central source,
but if angular momentum is conserved in the wind, $v_{\phi, \rm{wind}}
\sim 4$~km~s$^{-1}$ at $r = 20$~pc in this model.  This is entirely
negligible as the instrument resolution is $\sim 40$~km~s$^{-1}$.  The
presence of rotation does slightly modify the position of the critical
point; we have modified our 1D wind models to include rotation
\citep[and verified those models by reproducing, to within 0.1\%, the
results of][]{KG99} and found that rotation modifies the position of
the Parker wind's critical point by only 3\%; this is relatively
insignificant next to the 23\% change in the critical point caused by
the adoption of the non-point-source gravitational potential.  Thus,
for these models, we have neglected the effects of rotation.

\subsection{The Embedded Clouds: Their Origin, Longevity, and Fate}

The clouds in this model are chiefly used to try to slow the apparent
acceleration observed in \oiii: our main concern in this paper is to
attempt to reproduce the observed kinematics.  However, it is
interesting to ask about the origin of these clouds, their mass
outflow rate, and the various effects of the clouds embedded in the
thermal wind.  

First, a general overview of the clouds in this model.  It is
suggested by \citet{Das05} that STIS can resolve the cloud sizes, so
we consider a typical size of a cloud to be $\ga 1$pc.  For the ratio
of $n_{\rm wind}/n_{\rm cloud}$ found in this model, $\dot{M}_{\rm
cloud} \sim 24.5~C_{\rm f,cloud}~M_{\odot}$ yr$^{-1}$ where $C_{\rm
f,cloud}$ is of order $0.1~C_{\rm f,wind} \sim 0.02$.  This implies a
final $\dot{M}_{\rm cloud} \sim 0.5~M_{\odot}$ yr$^{-1}$.

\subsubsection{Cloud Origins}

Most importantly, in this model, for the parameters that fit this
data, clouds cannot be injected into the outflow at the launch point
of the wind.  The clouds must be injected downstream (the injection in
this wind happens at a radius of 20~pc), where the thermal wind's ram
pressure is sufficient to override the force of gravity felt by the
clouds.  If the clouds are injected into or exist in the model at
lower radii, those clouds fall back to the base of the wind.  

We do not constrain the origin of these clouds any further.  However,
it is reasonable to hypothesis that these clouds (given their size)
are portions of large-scale clouds in the inner regions of the galaxy
that are entrained in the hot thermal wind. One can also ask if the
embedded clouds may form from thermal instabilities in the wind, but
our Cloudy models of the wind are stable with the addition of
adiabatic cooling (without such cooling, Figs.~\ref{comptonTemp} and
\ref{altEquilCurve} show that such a wind at $T \sim 10^6$~K would be
thermally unstable, but those figures include only heating by the
central continuum and do not include adiabatic cooling).  In addition,
if clouds were created from thermal instabilities in within the wind,
they would share the velocity of the wind; the thermal wind itself
accelerates too quickly to explain the slow acceleration of the \oiii\
clouds, so cloud formation by thermal instability would not explain
the observations.

\subsubsection{Cloud Emissivity}

It is interesting to check whether the emissivity in the clouds is
approximately equal to the observed luminosity in \oiii.  In addition,
is it possible that the clouds, which must be limited in mass in order
to be dragged in the wind, cannot have enough mass to account for the
\oiii\ emission?

This is relatively easy to check.  \citet{Das05} report that the
\textsc{[O III]} flux from clouds is of order $10^{-12}$ to
$10^{-14}$~ergs cm$^{-2}$ s$^{-1}$; in particular, near the midpoint
of the clouds acceleration at 60~pc, the flux is $\sim 1.6 \times
10^{-14}$~ergs cm$^{-2}$ s$^{-1}$.  With NGC~4151 at a distance of
$\sim 13.3$~Mpc, this implies a flux of order $L_{\rm [O~III],
observed} \sim 3 \times 10^{38}$ergs~s.  The clouds in this model have
a total luminosity of $L_{\rm [O III], theory} = 4 \pi j_{\rm [O~III]}
V$ where $j_{\rm [O~III]}$ is the \oiii\ line emissivity (in units of
erg s$^{-1}$ cm$^{-3}$ ster$^{-1}$) and $V$ is the total emission
volume.  We estimate the volume as $\epsilon \frac{4}{3} \pi r^3$
where $\epsilon$ is the cloud volume filling factor, and $r_{\rm max}$
is the outer radius in the wind.  To estimate the maximum filling
factor for the clouds, we set $\dot{M}_{\rm wind}v_{\rm wind}C_{\rm
f,wind} \sim \dot{M}_{\rm wind}v_{\rm wind} C_{\rm f,max,clouds}$
where $C_{\rm f,max,clouds}$ is the maximum possible cloud covering
fraction.  For the mass outflow rates inferred in the wind, and the
cloud velocities in the wind, we find that $C_{\rm f,max,clouds} =
0.17$; this is a marginally smaller covering fraction than the wind
itself.  We then can estimate the cloud volume filling fraction,
$\epsilon \la 0.02$.  For clouds in the best-fitting wind, we can
estimate $j_{\rm [O~III]} \sim 5 \times 10^{-21}$ erg s$^{-1}$
cm$^{-3}$ ster$^{-1}$, which implies $L_{\rm [O~III], theory} \la 6
\times 10^{41}$~ergs~s$^{-1}$.  It is somewhat more reasonable to
assume that $\epsilon \sim 6 \times 10^{-4}$ \citep{Alexander99},
which implies $L_{\rm [O~III], theory} \sim 4 \times
10^{40}$~ergs~s$^{-1}$; for the simple order-of-magnitude check
employed here, this calculation shows that the clouds could indeed
account for the observed emissivity, and shows that the cloud filling
fraction is reasonably small.

\subsubsection{Cloud Longevity}

It is also interesting to check the Kelvin-Helmholtz timescales against
the acceleration timescales of the clouds to see whether these clouds
might be unstable to either Kelvin-Helmholtz or Rayleigh-Taylor
instabilities.  Both timescales are of order \citep[e.g.,][]{BdKS91}
\begin{eqnarray}
t_{\rm KH} & \sim & \frac{R_{\rm cloud}}{\delta v} \left( \frac{n_{\rm
    cloud}}{n_{\rm wind}} \right)^{1/2}  \sim  \left( \frac{R_{\rm cloud}
    N_{\rm cloud}}{n_{\rm wind}} \right)^{1/2} \frac{1}{\delta v} \\
& \sim & 2.2 \times 10^{19}~{\rm s} \left( \frac{R_{\rm cloud}}{1~{\rm pc}}
    \right)^{1/2} \frac{1}{\delta v}  
\end{eqnarray}

Meanwhile, the acceleration timescale is of order

\begin{equation}
t_{\rm acc} \sim \frac{\delta R}{v} \sim 3.1 \times 10^{11}~{\rm
  s}~\frac{R_{\rm pc}}{v_7}
\end{equation}
where $R_{\rm pc}$ is the acceleration length scale in parsecs, and
$v_7$ is the velocity in units of $10^7$~cm~s$^{-1}$.  For the
parameters of the clouds in our best-fit $n_0 = 2 \times
10^9$~cm$^{-3}$ model, $t_{\rm KH} > t_{\rm acc}$ for the clouds out
to $r \sim 30$~pc, and $t_{\rm KH}$ is less than $t_{\rm acc}$ by only
a factor of less than three, so only large clouds (on scales of 6~pc
or more, which would be approximately the size of clouds that Das et
al. seem to be resolving in NGC~4151, where 6~pc $\sim 0.1''$).  Our
chief concern in this model is in matching the observed dynamics,
however.  

We can also investigate whether these clouds would be confined by the
thermal pressure of the wind.  Interestingly, although we did not
attempt to satisfy this constraint in our models, the final cloud
models are close to being in thermal pressure equilibrium with the
thermally-driven wind.  We do not model the thermal equilibrium of the
clouds, but assuming a cloud temperature of $T_{\rm cloud} \sim
10^4$~K and a fiducial wind temperature of $T_{\rm wind} \sim 2 \times
10^6$~K, clouds of radius $\sim 0.2$~pc would be in approximate
thermal equilibrium.  Clouds of larger size, such as the clouds
apparently required here, would have proportionately smaller densities
in our model (in order to reproduce the velocity curves), and so have
lower pressures for the assumed $T_{\rm cloud}$.

\begin{figure}[h]
\begin{center}
\includegraphics[angle=-90, width=8cm]{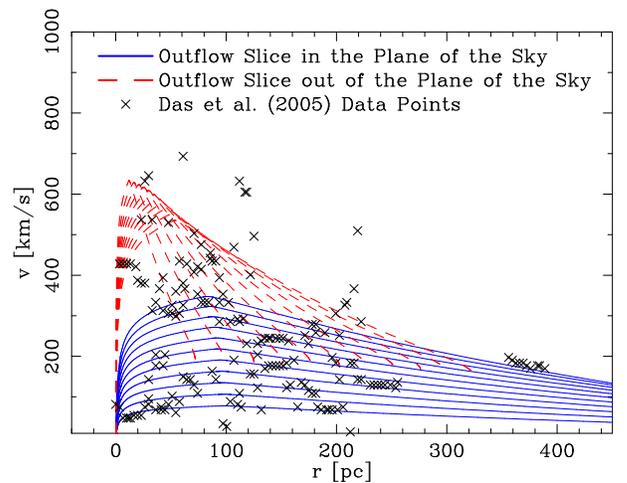}
\caption{The velocity vs distance profiles of the biconical outflow
  for the best fit model for the wind alone, without any embedded
  clouds.  The data points are the same as those in
  Figures~\ref{cloudModelBestFit} and
  \ref{windMotionForCloudModelBestFit}. \label{windBestFit}}
\end{center}
\end{figure}

We note that it is possible to find a Parker wind solution without
clouds that also fits the data, although not as well as the wind and
cloud model; note in particular the data points that are not fit near
($r$,$v$) values of (20~pc, $v \sim 35$~km~s$^{-1}$), (20~pc,
400~km~s$^{-1}$), as well as those points above the $r > 100$~pc
velocity envelope of the wind.  This solution is displayed in
Figure~\ref{windBestFit}, with parameters listed in
Table~\ref{windResults}; note that we fit the wind-only model using
the same geometry as used in the wind+cloud model.  Since this
wind-only model simultaneously results in a somewhat worse fit and
similar parameters for the previous model, we consider the previous
wind and cloud model the best representation of the data.
\begin{deluxetable*}{lll}
\tablecaption{Inferred Parameters for the NLR in NGC 4151 Using Wind Model\label{windResults}} 
\tablehead{
\colhead{Quantity} & \colhead{Value} & \colhead {Range Permissible for
Reasonable Fit}}
\startdata
Wind Temperature, $T$                &  $2.2 \times 10^6$~K & $2.0 \times 10^6 - 2.4 \times 10^6$ \\
Launch Radius, $r_{\rm launch}$      &  $0.1$~pc &  $< 0.8$~pc \\
ISM Gas Radius, $r_{\rm ISM}$        &  $100$~pc & $50 - 210$~pc \\
$\frac{n_{\rm wind}}{N_{\rm H,ISM~cloud}}$ & $1.1 \times
10^{-21}$~cm$^{-1}$ & $7.4 \times 10^{-22} - 1.4 \times 10^{-21}$~cm$^{-1}$ \\
Angle of Inclination of Outflow Cone & $45^{\circ}$ & \\
Inner Cone Opening Angle             & $12^{\circ}$ & \\
Outer Cone Opening Angle             & $38^{\circ}$ & \\
\enddata
\end{deluxetable*}
\section{Tests of Photoionization Consistency: Can an
  isothermal wind exist in NGC 4151?}\label{testingIsothermality}

Having defined an optimum temperature to fit the observations (see
Table~\ref{cloudResults}), we now test whether the temperatures
required in the isothermal models can be achieved via photoionization
from the central AGN.  We also check for how the temperature varies as
a function of radius in the wind, to test if the wind does indeed have
$T(r)$ decreasing less quickly than $1/r$ on small scales and $T(r)$
at least constant on large scales, as required by the Parker wind
model (see \S\ref{OverviewParkerWind}).  Note that for the initial
photoionization tests we present, the run of density vs. radius from
the isothermal Parker-wind model is used to set the density at each
radius in the wind, so these models are not completely
self-consistent.  We will then conclude by testing an ``X-ray rich''
continuum to see if this can help explain the observations, and then
finally we will test a non-isothermal Parker wind (with a
photoionization-derived $T(r)$ profile) against the observations.

\subsection{Photoionization Results}\label{photoResults}

In order to look for a physically consistent thermal wind model that
matches the earlier isothermal model and the observations, we ask (1)
whether a roughly isothermal temperature structure can be achieved in
an AGN wind, (2) whether the temperature is high enough to roughly
match the temperature required in the isothermal wind model, and (3)
whether $T(r)$ decreases less quickly than $M(r)/r$.

\begin{figure}[h]
\begin{center}
\includegraphics[angle=-90, width=8cm]{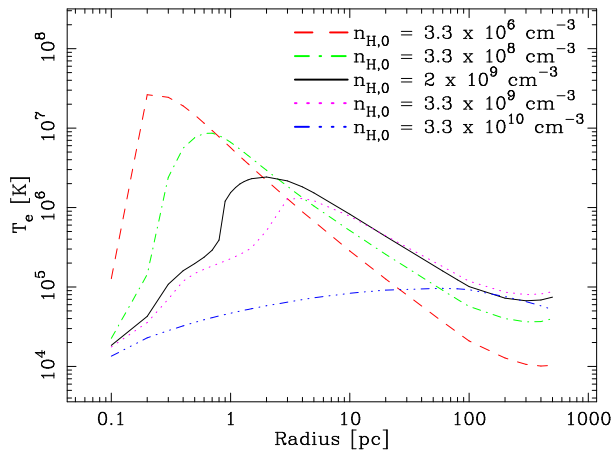}
\caption{$T_{\rm e}$ in the wind as a function of radius in the Parker
  wind.  As shown in the figure, both the rise in $T_{\rm e}$ as a
  function of radius and the maximum $T_{\rm e}$ achieved are strongly
  dependent on the initial density.  Only winds with $n_{\rm H,0} \la
  2 \times 10^9$~cm$^{-3}$ achieve temperatures of $T \sim 3 \times
  10^6$~K required by the isothermal Parker wind model.  Note also the
  slower increase of $T_{\rm e}$ with radius as the density increases.
  \label{temperatureStructure}}
\end{center}
\end{figure}

To answer the first two questions, we calculate the temperature as a
function of radius for winds with a range of initial densities,
$n_{\rm H,0}$, displayed in Figure~\ref{temperatureStructure}.  The
temperature vs. radius for each $n_{\rm H,0}$ is calculated by taking
the $n(r)$ profile from the isothermal wind model and running Cloudy
photoionization simulations at a range of radii given the central
continuum of NGC~4151 (defined in \S\ref{sed}).  This figure shows
that only for very large $n_{\rm H,0}$ is the wind approximately
isothermal.  The only initial density in
Figure~\ref{temperatureStructure} that leads to an approximately
isothermal wind is $n_{\rm H,0} = 3.3 \times 10^{10}$~cm$^{-3}$.  This
density is so high that the resultant wind temperatures rise only to
$T_{\rm e,max} \sim 10^5$~K.  Such a temperature is far too low to
explain the observed velocities in NGC~4151 which required $T_{\rm e}
= 3 \times 10^6$~K.  Lower densities ($n_{\rm H,0} \la 3.3 \times
10^9$) are also not possible in this model as $T(r)$ always drops
significantly at large distances.  This is simply due to adiabatic
cooling.  With adiabatic cooling switched off, the temperature in the
wind very quickly achieves $T_e \approx 3 \times 10^7$~K and remains
at that constant temperature over the rest of the outflow.  Even with
adiabatic cooling included, however, Figure~\ref{temperatureStructure}
also shows that $n_{\rm H,0} \sim 2 \times 10^9$~cm$^{-3}$ can achieve
approximately the required temperatures over at least a limited range
of radii in the outflow; the drop in $T(r)$ at larger distances shows
that it may still be problematic, however.

\begin{figure}[h]
\begin{center}
\includegraphics[angle=-90, width=8cm]{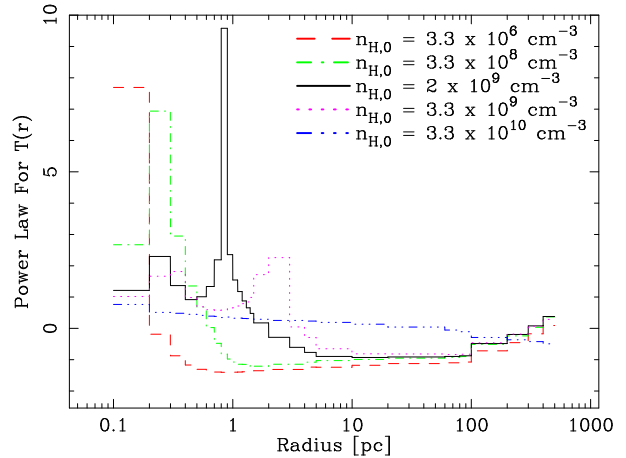}
\caption{A piecewise power-law fit to the temperature data in
  Figure~\ref{temperatureStructure}, with the power law, $\alpha$
  defined via $T \propto r^{\alpha}$ for radii between the
  photoionization simulation results.  Physically consistent Parker
  winds in the NLR of NGC~4151 would require $T(r)$ dropping less
  quickly than $1/r$ out to $r \sim 2$~pc, and $T(r)$ at least
  constant out to $r \sim 10$~pc in order for the wind to accelerate.
  As can be seen here, this is very difficult for any of the winds
  (given the dominance of adiabatic cooling at large radii) to
  satisfy.\label{temperaturePowerLaw}}
\end{center}
\end{figure}

Next, we focus on how quickly the temperature can change with radius
and still allow a Parker wind to form.
Figure~\ref{temperaturePowerLaw} displays the exponent of a piecewise
power law that fits the temperature profile shown in
Figure~\ref{temperatureStructure}.  Again, in order for a physically
consistent Parker wind to exist in the core of NGC~4151, $T(r)$ must
drop less slowly than $1/r$ for small distances ($r \la 2$~pc), and
then, where $M(r)$ increases almost linearly with radius, $T(r)$ must
be, at least, approximately constant: it must have a power law index
$> 0$.  This is true for the high density run with $n_{\rm H,0} = 3.3
\times 10^{10}$~cm$^{-3}$, but as already shown in
Figure~\ref{temperatureStructure}, that wind does not have
temperatures high enough to yield the observed velocities.  The lower
density trials meanwhile show extraordinarily large changes in $T_{\rm
e}$ with radius.  For $n \le 3.3 \times 10^8$~cm$^{-3}$, $T(r)$ drops
below a power law index value of zero at $r < 0.7$~pc.  In contrast,
acceleration of the \textit{isothermal} Parker wind must continue out
to a few times the Parker wind's critical point distance, which is at
$r_{\rm crit} \sim 1$~pc for the isothermal wind model shown in
Figure~\ref{cloudModelBestFit} (non-isothermal wind models have
critical points even further out, since much of the wind's
acceleration occurs with $T(r) < T_{\rm isothermal}$).  However, the
temperature is already dropping precipitously at radii even smaller
than $r_{\rm crit}$ for $n_{\rm H,0} \le 3.3 \times 10^8$~cm$^{-3}$,
so that range of densities would not yield a Parker wind.

\begin{figure}[h]
\begin{center}
\includegraphics[angle=-90, width=8cm]{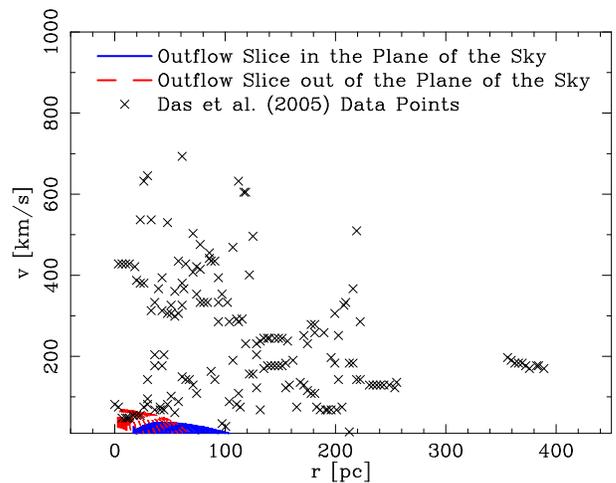}
\caption{As in Figure~\ref{cloudModelBestFit}, but with a
  Cloudy-derived $T(r)$ profile for $n_{\rm H,0} = 2 \times
  10^9$~cm$^{-3}$.  The wind, with a maximum velocity of only
  $\sim$365 km~s$^{-1}$ (the cloud velocities, shown here, do not even
  approach that velocity), and decelerating for $r > 8$~pc, cannot
  drag the clouds to anywhere near the observed
  velocities.\label{variableTr-n2e9}}
\end{center}
\end{figure}

The test case of $n_{\rm H,0} = 2 \times 10^9$~cm$^{-3}$ is closer to
having the required temperature (at least at its maximum temperature),
and is closer to the desired $T(r)$ profile, and so more careful
consideration is justified in this case.  As shown in
Figure~\ref{temperatureStructure}, this model does, if only in a
limited range of radii, achieve approximately the temperature of the
best-fitting isothermal model.  Figure~\ref{temperaturePowerLaw} shows
that the temperature does not drop appreciably until $r \sim 2$~pc.
We would expect that since acceleration in the wind is required out to
a few parsecs, this temperature profile would also not lead to a
viable Parker wind model.  But, as the temperature profile is close to
satisfying the requirements on $T(r)$, we build a numerical Parker
wind model for generic $T(r)$ from our photoionization simulations
that finds, self-consistently, the critical point for that particular
temperature law.  Applying this model to $T(r)$ for the case of
$n_{\rm H,0} = 2 \times 10^9$~cm$^{-3}$, we found that the wind
decelerates beyond $r = 8$~pc, and so would not fit the required
$v(r)$ profile.  Testing to see if this $n_{\rm H,0} = 2 \times
10^9$~cm$^{-3}$ profile results in clouds velocities that might fit
the data, we apply the same geometry and drag parameters as used in
the previous best-fit model (see Table~\ref{cloudResults}) and present
the model vs. the data from \citet{Das05} in
Figure~\ref{variableTr-n2e9}.  The model clearly cannot fit the
observed velocities, or their variation with distance.

\begin{figure}[h]
\begin{center}
\includegraphics[angle=-90, width=8cm]{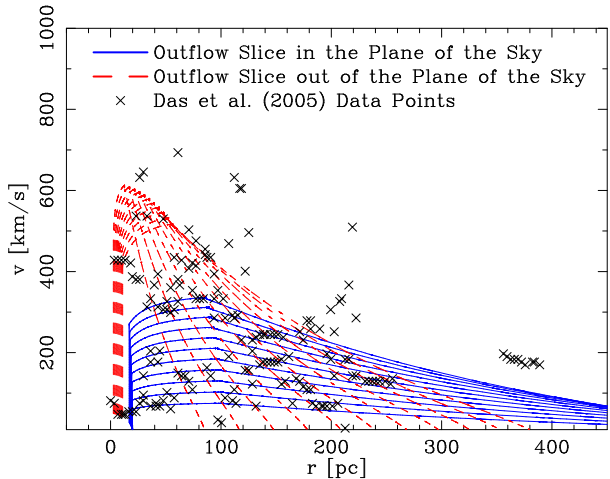}
\caption{As in Figure~\ref{cloudModelBestFit}, but with a
  Cloudy-derived $T(r)$ profile for $n_{\rm H,0} = 3.3 \times
  10^8$~cm$^{-3}$.  \label{variableTr-n3e8}}
\end{center}
\end{figure}

If we instead use the $T(r)$ profile derived from the $n_{\rm H,0} =
3.3 \times 10^8$~cm$^{-3}$ wind model, we again find a wind that
decelerates too quickly (here, the wind starts decelerating at $r =
4$~pc).  Attempting to fit this model to the data from \citet{Das05}
results in a much better fit (see Figure~\ref{variableTr-n3e8}) than
in Figure~\ref{variableTr-n2e9}, but still misses the high velocity
data points shown in the figure.  Modifying the drag term in our wind
equation does not improve the fit: decreasing the drag term to fit the
higher-velocity points simply increases the outflow velocities,
therefore doing an increasingly worse job of fitting the data points
near $(r,v) = (70~{\rm pc}, 350~{\rm km~s}^{-1})$.  Also, modifying
the geometry of the outflow has no significant impact on the fit: the
half-angle of the bicone is already $38^\circ$ here; increasing that
to a maximum value of $45^\circ$ helps to include some of the higher
velocity data points, but cannot make up for the discrepancy relative
to the isothermal model.  Still, this model represents the closest fit
to the data achievable with a more realistic $T(r)$ profile; there is
no quantitative way to distinguish this fit from the previous
isothermal best-fit, but it is perhaps encouraging; this model will be
examined in more detail in \S\ref{selfconsistantTr}.  Unfortunately,
all of the rest of the generic $T(r)$ models cannot come close to
reproducing the data (as can be seen, for example, in
Fig.~\ref{variableTr-n2e9}).

\subsection{How does a self-consistent $T(r)$ affect the Parker wind
  simulations?}\label{selfconsistantTr} 

Up until this point, we have used the $n(r)$ profile from an
isothermal Parker wind model as input to a set of photoionization
models, only rescaling the base density to test different initial
densities.  It is important to check if the fitting results improve
with a more self-consistent $T(r)$, $n(r)$, and $v(r)$ profile instead
of using the isothermally-derived $n(r)$ profile to calculate $T(r)$,
as before.

We test this possibility by examining the $T(r)$ profile from the $n_0
= 3.3 \times 10^8~{\rm cm}^3$ simulation (shown in
Fig.~\ref{temperatureStructure}).  This is then input into our
non-isothermal Parker wind model, and the resulting acceleration
profile and $n(r)$ profile is computed.  This $n(r)$ profile is then
used as the basis for another iteration of Cloudy simulations, to
compute the resultant $T(r)$.  We ask whether the result of this
iteration on the $T(r)$ structure will result in a temperature profile
more suitable for a large-scale Parker wind.  However, the resulting
$T(r)$ profile only results in a worse fit to the data: the rise in
the initial $T(r)$ profile at $r < 1$~pc causes the wind model to
accelerate from the disk faster than the fiducial isothermal wind
model.  This increase in acceleration yields a wind that must be
launched from larger radii ($r \sim 0.2$~pc instead of $0.1$~pc) and
also has a faster dropoff in density with height, which then yields a
$T(r)$ curve closer to the lower-density curves in
Figure~\ref{temperatureStructure}: the self-consistent $T(r)$ only
results in a less isothermal wind model whose temperature decreases
more quickly with radius beyond $r \approx 1$~pc.  Therefore,
attempting to include a self-consistent temperature profile in the
Parker wind model only results in a more rapidly increasing (and then
more rapidly decreasing) $T(r)$ profile, which results in a wind model
that has a $T(r)$ profile even less suited for accelerating a wind
that might fit the observations of \citet{Das05}.  So, in the end,
including such a $T(r)$ profile in the Parker wind also cannot explain
the observed acceleration in NGC~4151.

\subsection{Does an X-ray ``rich'' continuum help?}\label{altContinuum}

\begin{figure}[h]
\begin{center}
\includegraphics[angle=-90, width=8cm]{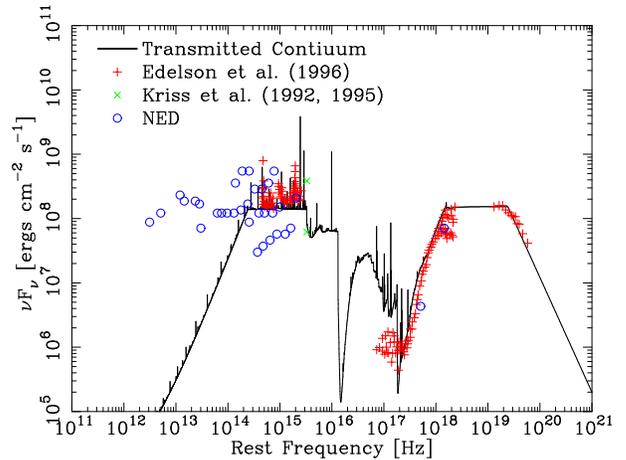}
\caption{An alternate X-ray ``rich'' incident continuum for NGC~4151,
  plotted with the same data as in Fig.~\ref{sedCompare}.  For this
  continuum, we explicitly set the high-energy X-ray continuum to fit
  \citet{Edelson96}, as opposed to the previously adopted continuum, which
  adopted the \citet{Kraemer00} power law in the X-rays.  This led to
  slightly higher X-ray flux.  At the same time, the continuum below 1
  eV was suppressed to lower Compton cooling in the outflow \citep[as
  in][]{Krolik99}.  \label{altSED}}
\end{center}
\end{figure}

To test the limits of our model of the central continuum, we now adopt
a relatively X-ray-bright continuum to check if such a continuum would
heat the gas to higher temperatures and modify our results.  Displayed
in Figure~\ref{altSED}, this continuum was primarily modified in the
high-energy X-ray regime to more closely follow the observations of
\citet{Edelson96} and the inferred high-energy spectrum of
\citet{Alexander99}; we continue to include the UV absorption, as
before.  This increase in X-ray brightness increases Compton heating
within the gas.  Meanwhile, the low-energy continuum in the SED has
been significantly cut off below 1~eV in order to reduce Compton
cooling by low-energy photons as much as possible.  This was done, for
instance, in \citet{Krolik99} to account for the possibility that
lower-energy photons are emitted at much larger scales and so might
not act to cool the outflow in the immediate vicinity of the central
black hole.  (In reality, for the large-scale winds tested here, the
question of the origin of the low-energy photons is less important
since they almost certainly play a role on scales of 10~pc.  However,
to test the limits of the input continuum model, we attempt such a
continuum anyway.)

\begin{figure}[h]
\begin{center}
\includegraphics[angle=-90, width=8cm]{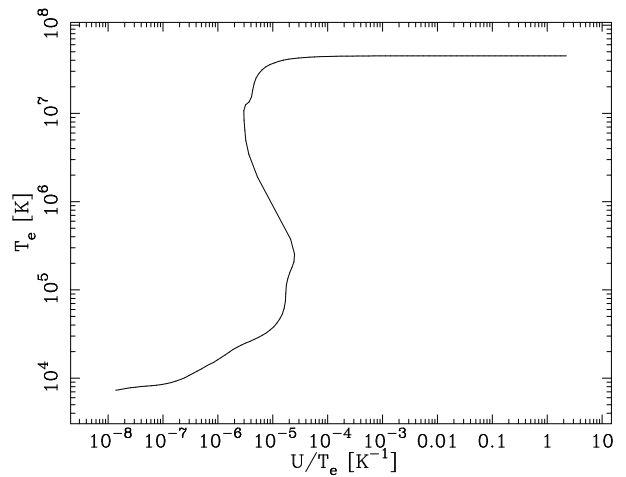}
\caption{The photoionization equilibrium curve for the SED shown in
  Fig.~\ref{altSED}.  As can be seen by comparing to
  Fig.~\ref{comptonTemp}, $T_{\rm Compton}$ does not change
  appreciably.  \label{altEquilCurve}}
\end{center}
\end{figure}

Adopting this continuum does not greatly modify the Compton
temperature of the gas.  The radiative equilibrium curve for this
continuum is shown in Figure~\ref{altEquilCurve}; for this continuum,
$T_{\rm Compton} = 4.2 \times 10^7$~K as opposed to $T_{\rm Compton} =
3 \times 10^7$~K for the original continuum.

As one might infer from that relatively small change in the Compton
temperature of the gas, adopting this continuum for calculating $T(r)$
does not result in significant changes to our results.  We also
calculated a self-consistent Parker wind after inputing this continuum
into our photoionization simulations, which yielded no appreciable
difference from the tests with the original SED.  Overall, adopting
this SED does not significantly alter any of our conclusions.  An
X-ray ``rich'' continuum for the central SED of NGC~4151 cannot help a
Parker wind model to explain the observed velocity trends.

\section{Conclusions}

We have developed models of Parker winds accounting for the effects of
the radially varying $M(r)$, adiabatic cooling, and drag effects
between the wind and embedded clouds as well as drag forces between
the wind and an external medium.  We then tested these models against
the conditions that exist in the center of one particular local AGN,
NGC~4151.  We have shown that simple thermal winds cannot explain the
slow acceleration that has been inferred for the NLR of NGC~4151.  The
strongest constraint against self-consistent Parker winds is that
adiabatic cooling on the large scales of these outflows leads to
temperatures in the wind that decreases strongly as a function of $r$.
With such a temperature profile, Parker winds cannot be launched and
accelerated to the observed velocities at the observed distances.
Varying the initial density in the wind either yields temperatures
that drop quickly with distance (for $n_{\rm H,0} \la 10^8$~cm$^{-3}$)
or temperatures that are too low to reproduce the observed velocities
($n_{\rm H,0} \ga 10^{9}$~cm$^{-3}$).  The intermediate case of
$n_{\rm H,0} \sim 3 \times 10^8$~cm$^{-3}$ comes closest to
reproducing the data, yielding a temperature profile that we have
checked by developing a numerical Parker wind model for general
$T(r)$.  For the temperature profile shown in
Figure~\ref{temperatureStructure}, the wind starts decelerating
already at $r \sim 4$~pc; a radius much smaller than the 100~pc out to
which acceleration is inferred in NGC~4151.  We find that plotting the
resultant predicted $v(r)$ against the data from \citet{Das05} shows
that this wind model cannot reproduce the range of \oiii\ velocity
points (see Fig.~\ref{variableTr-n3e8}) as well as the isothermal wind
model (which cannot be obtained for NGC~4151).  Further, attempting to
develop a self-consistent $T(r)$ profile for this wind results in a
temperature profile even further from isothermal which correspondingly
fails to fits the data.  In short, we find no self-consistent $T(r)$
profile for a large-scale Parker wind that can explain the observed
ranges of velocities.

If thermal winds are not driving the observed $v(r)$ profile, how can
it be explained?  Radiative driving is most likely not the source of
the acceleration: we have tested simple models of radiative
acceleration on dust, and such models achieve terminal velocities at
distances of order the launch radius -- much closer to the launch
point than the observed velocity profiles.  Magnetic winds would also
achieve their terminal velocities on similar scales (although perhaps
clouds dragged in the magnetic wind might yield the slow acceleration
seen).

The difficulty in fitting this data with almost any known wind model
leads us to question whether the observed velocity law is actually a
wind accelerating with radius.  Perhaps the observed kinematic profile
is the result of an interaction of a wind with the surrounding medium.
An interesting alternative interpretation is suggested in the work of
\citet{MM99}, which showed that rather generic hydromagnetic winds,
expanding into gas surrounding a young stellar object at the wind's
terminal velocity, would yield $v \propto r$, or a ``Hubble law''-type
flow.  This is a promising theory for the accelerating portion of the
outflow, but it seems unclear how the observed slow deceleration of
the flow in NGC~4151 could then be explained in the same picture.
Still, it is important to point out that given the inability of a wide
range of wind models to explain the observed $v(r)$ profile, perhaps
these observations are not revealing acceleration of a wind, but
instead the interaction of an already accelerated wind with the
surrounding medium.

\section{Acknowledgments}
The authors wish to thank Mike Crenshaw, Alvin Das, Steve Kraemer and
Chris Matzner for helpful conversations while this work was in
progress.  We also thank our referee for a helpful and insightful
report.  Many thanks to Gary Ferland and his collaborators for the
development, distribution, and support of Cloudy, which was used in
this work.  This research was supported by the Natural Sciences and
Engineering Research Council of Canada.  JEE was also supported by NSF
AST-0507367 and NSF PHY-0215581 (to the Center for Magnetic
Self-Organization in Laboratory and Astrophysical Plasmas).  This
research has made use of the NASA/IPAC Extragalactic Database (NED)
which is operated by the Jet Propulsion Laboratory, California
Institute of Technology, under contract with the National Aeronautics
and Space Administration.  This research has made use of NASA's
Astrophysics Data System.


\begin{thebibliography}{}
\bibitem[Alexander et al.(1999)]{Alexander99}Alexander, T., Sturm, E.,
  Lutz, D., Sternberg, A., Netzer, H. \& Genzel, R. 1999, ApJ, 512,
  204
\bibitem[Arav, Li \& Begelman(1994)]{ALB94}Arav, N., Li, Z.-Y.,
  Begelman, M. 1994, ApJ, 432, 62
\bibitem[Axon et al.(1998)]{Axon98}Axon, D.J., Marconi, A., Capetti,
  A., Machetto, F.D., Schreier, E. \& Robinson, A. 1998, ApJ, 496, 75 
\bibitem[Balsara \& Krolik(1993)]{BK93}Balsara, D.S. \& Krolik
  J.H. 1993, ApJ, 402, 109
\bibitem[Begelman, deKool \& Sikora(1991)]{BdKS91}Begelman, M.C., de
  Kool, M. \& Sikora, M. 1991, ApJ, 382, 416
\bibitem[Begelman, McKee \& Shields(1983)]{BMS83}Begelman, M.C.,
  McKee, C.F., Shields, G.A. 1983, ApJ, 271, 70
\bibitem[Blandford(1973)]{B73}Blandford, R.D. 1973, A\&A, 26, 161
\bibitem[Blandford \& Payne(1982)]{BP82}Blandford, R.D. \& Payne,
  D.G. 1982, MNRAS, 199, 883
\bibitem[Bondi \& Hoyle(1944)]{BH44}Bondi, H. \& Hoyle, F., 1944,
  MNRAS, 104, 273
\bibitem[Bottorff et al.(1997)]{Bottorff97}Bottorff, M., Korista,
  K.T., Shlosman, I., Blandford, R.D. 1997, ApJ, 479, 200
\bibitem[Bottorff \& Ferland(2000)]{BF00}Bottorff, M.C. \& Ferland,
  G.J. 2000, MNRAS, 316, 103
\bibitem[Bottorff, Korista \& Shlosman(2000)]{Bottorff00}Bottorff,
  M.C., Korista, K.T. \& Shlosman, I. 2000, ApJ, 537, 134
\bibitem[Brighenti \& Mathews(2006)]{BM06}Brighenti, F. \& Mathews,
  W.G. 2006, ApJ, 643, 120
\bibitem[Capetti, Macchetto \& Lattanzi(1997)]{Capetti97}Capetti, A.,
  Macchetto, F.D., Lattanzi, M.G. 1997, ApJ, 476, L67
\bibitem[Chelouche \& Netzer(2001)]{CN01}Chelouche, D. \& Netzer,
  H. 2001, MNRAS, 326, 916
\bibitem[Chelouche \& Netzer(2003a)]{CN03a}Chelouche, D. \& Netzer,
  H. 2003, MNRAS, 344, 223
\bibitem[Chelouche \& Netzer(2003b)]{CN03b}Chelouche, D. \& Netzer,
  H. 2003, MNRAS, 344, 233
\bibitem[Chelouche \& Netzer(2005)]{CN05}Chelouche, D. \& Netzer,
  H. 2005, ApJ, 625, 95
\bibitem[Chiang \& Murray(1996)]{CM96}Chiang, J. \& Murray, N. 1996,
  ApJ, 466, 704
\bibitem[Contopolous \& Lovelace(1994)]{CL94}Contopoulos, J. \&
  Lovelace, R.V.E. 1994, ApJ, 429, 139
\bibitem[Crenshaw et al.(1999)]{Crenshaw99}Crenshaw, D.M., Kraemer,
  S,B., Boggess, A., Maran, S.P., Mushotzky, R.F., \& Wu, C.-C. 1999,
  ApJ, 516, 750
\bibitem[Crenshaw \& Kraemer(2000)]{CK00}Crenshaw, D.M. \& Kraemer,
  S.B., 2000, ApJ, 532, L101
\bibitem[Dalgarno \& McCray(1972)]{DM72}Dalgarno, A. \& McCray,
  R. A. 1972, ARA\&A, 10, 375
\bibitem[Das et al.(2005)]{Das05}Das, V., et al. 2005, AJ, 130, 945
\bibitem[Edelson et al.(1996)]{Edelson96}Edelson, R.A. et al. 1996, ApJ,
  470, 364
\bibitem[Emmering, Blandford \& Shlosman(1992)]{EBS92}Emmering, R.T.,
  Blandford, R.D. \&Shlosman, I. 1992, ApJ, 385, 460 
\bibitem[Everett, K\"onigl, \& Arav(2002)]{EKA02}Everett, J.E.,
  K\"onigl, A. \& Arav, N. 2002, ApJ, 569, 671
\bibitem[Everett(2005)]{E05}Everett, J.E. 2005, ApJ, 631, 689
\bibitem[Ferland et al.(1998)]{Ferland98}Ferland, G. J. Korista,
K.T. Verner, D.A. Ferguson, J.W. Kingdon, J.B. Verner, \& E.M. 1998,
PASP, 110, 761
\bibitem[Frank, Noriega-Crespo \& Balick(1992)]{Frank92}Frank, A.,
  Noriega-Crespo, A. \& Balick, B. 1992, AJ, 104, 841
\bibitem[George et al.(1998)]{George98}George, I.M., Turner, T.J.,
  Netzer, H., Nandra, K., Mushotzky, R.F. \& Yaqoob, T. 1998, ApJS,
  114, 73
\bibitem[Goodman(2003)]{Goodman03}Goodman, J. 2003, MNRAS, 339, 937
\bibitem[Icke(1977)]{Icke77}Icke, V. 1977, Nature, 266, 699
\bibitem[Keppens \& Goedbloed(1999)]{KG99}Keppens, R. \& Goedbloed,
  J.P. 1999, A\&A, 343, 251
\bibitem[K\"onigl \& Kartje(1994)]{KK94}K\"onigl, A. \& Kartje,
  J.F. 1994, ApJ, 434, 446
\bibitem[de Kool \& Begelman(1995)]{dKB95}de Kool, M. \& Begelman,
  M.C. 1995, ApJ, 455, 448
\bibitem[Kraemer \& Crenshaw(2000)]{KC00}Kraemer, S.B. \& Crenshaw,
  D.M. 2000, ApJ, 544, 763
\bibitem[Kraemer et al.(2000)]{Kraemer00}Kraemer, S.B., Crenshaw,
  D.M., Hutchings, J.B., Gull, T.R., Kaiser, M.E., Nelson, C.H., \&
  Weistrop, D. 2000, ApJ, 531, 278
\bibitem[Kraemer et al.(2001)]{Kraemer01}Kraemer, S.B., et al., 2001,
  ApJ, 551, 671
\bibitem[Kraemer et al.(2005)]{Kraemer05}Kraemer, S.B., et al., 2005,
 ApJ, 633, 693
\bibitem[Kriss et al.(1992)]{Kriss92}Kriss, G.A. et al. 1992, ApJ, 392,
  485
\bibitem[Kriss et al.(1996)]{Kriss96}Kriss, G.A., Davidsen, A.F., Zheng,
W., Kruk, J.W. \& Espey, B.R. 1995, ApJ, 454, L7
\bibitem[Kriss(2001)]{Kriss01}Kriss, G.A. 2001, in Mass Outflow in
  Active Galactic Nuclei: New Perspectives, Eds. D.M. Crenshaw,
  S.B. Kraemer \& I.M. George (ASP: San Francisco), 255, 69
\bibitem[Krolik(1997)]{Krolik97}Krolik, J.H. 1997, Ap\&SS, 248, 207
\bibitem[Krolik(1999)]{Krolik99}Krolik, J.H. 1999, Active Galactic
  Nuclei: From the Central Black Hole to the Galactic Environment
  (Princeton: Princeton University Press)
\bibitem[Krolik \& Kriss(1995)]{KK95}Krolik, J.H. \& Kriss, G.A. 1995,
  ApJ, 447, 512
\bibitem[Krolik, McKee \& Tarter(1981)]{KMT81}Krolik, J.H., McKee,
  C.F. \& Tarter, C.B. 1981, ApJ, 249, 422
\bibitem[Krolik \& Vrtilek(1984)]{KV84}Krolik, J.H. \& Vrtilek,
  J.M. 1984, ApJ, 279, 521
\bibitem[Levich \& Sunyaev(1970)]{LS70}Levich, E.V. \& Sunyaev,
  R.A. 1970, Astroph. Lett., 7, 69
\bibitem[Matzner \& McKee(1999)]{MM99}Matzner, C.D. \& McKee,
  C.F. 1999, ApJ, 526, L109
\bibitem[McKee \& Tarter(1975)]{MT75}McKee, C.F. \& Tarter, C.B. 1975,
  ApJ, 202, 306
\bibitem[Moore, Cohen \& Marcy(1996)]{MCM96}Moore, D., Cohen, R.D.,
  Marcy, G.W. 1996, ApJ, 470, 280
\bibitem[Moore \& Cohen(1996)]{MC96}Moore, D. \& Cohen, R.D. 1996,
  ApJ, 470, 301
\bibitem[Mundell et al.(1999)]{Mundell99}Mundell, C.G., Pedlar, A.,
  Shone, D.L. \& Robinson, A. 1999, MNRAS, 304, 481
\bibitem[Murray et al.(1995)]{MCGV95}Murray, N., Chiang, J., Grossman,
  S.A., Voit, G.M. 1995, ApJ, 451, 498
\bibitem[Nelson \& Whittle(1996)]{NW96}Nelson, C.H. \& Whittle,
  M. 1996, ApJ, 465, 96
\bibitem[Nelson et al.(2000)]{Nelson00}Nelson, C.H. et al. 2000, ApJ,
  531, 257
\bibitem[Parker(1965)]{Parker65}Parker, E.N. 1965, SSRv, 4, 666
\bibitem[Peterson(1997)]{Peterson97}Peterson, B.M. 1997, An
  Introduction to Active Galactic Nuclei (New York:Cambridge
  University Press)
\bibitem[Peterson(2004)]{Peterson04}Peterson, B.M., et al. 2004, ApJ,
  613, 682
\bibitem[Proga \& Kallman(2004)]{PK04}Proga, D. \& Kallman, T.R. 2004,
  ApJ, 616, 688
\bibitem[Proga, Stone \& Kallman(2000)]{Proga00}Proga, D., Stone,
  J.M. \& Kallman, T.R., ApJ, 543, 686
\bibitem[Reynolds et al.(1997)]{Reynolds97}Reynolds, C.S. 1997, MNRAS,
  286, 513
\bibitem[Rice et al.(2006)]{Rice06}Rice, M.S., Martini, P., Greene,
  J.F., Pogge, R.W., Shields, J.C., Mulchaey, J.S. \& Regan,
  M.W. 2006, ApJ, 636, 654
\bibitem[Ruden, Glassgold \& Shu(1990)]{RGS90}Ruden, S.P.,
  Glassgold, A.E. \& Shu, F.H. 1990, ApJ, 361, 546
\bibitem[Ruiz et al.(2001)]{Ruiz01}Ruiz, J.R., Crenshaw, D.M.,
  Kraemer, S.B., Bower, G.A., Gull, T.R., Hutchings, J.B., Kaiser,
  M.E. \& Weistrop, D., 2001, AJ, 122, 2961
\bibitem[Ruiz et al.(2005)]{Ruiz05}Ruiz, J.R., Crenshaw, D.M.,
  Kraemer, S.B., Bower, G.A., Gull, T.R., Hutchings, J.B., Kaiser,
  M.E., Weistrop, D. 2005, AJ, 129, 73
\bibitem[Safier(1993)]{Safier93}Safier, P.N. 1993, ApJ, 408, 115
\bibitem[Schulz(1990)]{Schulz90}Schulz, H. 1990, AJ, 99, 1442
\bibitem[Shang et al.(2002)]{Shang02}Shang, H., Glassgold, A.E., Shu,
  F.H. \& Lizano, S. 2002, ApJ, 564, 853
\bibitem[Shields(1977)]{Shields77}Shields, G.A. 1977, Ap. Letters, 18, 119
\bibitem[Shlosman, Vitello \& Shaviv(1985)]{SVS85}Shlosman, I.,
  Vitello, P.A. \& Shaviv, G. 1985, ApJ, 294, 96
\bibitem[Smith(1984)]{Smith84}Smith, M.D. 1984, MNRAS, 209, 913
\bibitem[Steffen et al.(1997)]{Steffen97}Steffen, W., G\'omez, J.L.,
  Williams, R.J.R., Raga, A.C. \& Pedlar, A. 1997, MNRAS, 286, 1032
\bibitem[Wilson(1982)]{Wilson82}Wilson, D.B. 1982, MNRAS, 200, 881
\bibitem[Winge et al.(1999)]{Winge99}Winge, C. Axon, D.J., Macchetto,
  F.D., Capetti, A. \& Marconi, A. 1999, ApJ, 519, 134
\bibitem[Woods et al.(1996)]{Woods96}Woods, D.T., Klein, R.I., Castor,
  J.I., McKee, C.F., Bell, J.B. 1996, ApJ, 461, 767
\end{thebibliography}
\end{document}